\documentclass[conference]{IEEEtran}
\IEEEoverridecommandlockouts
\usepackage{acronym}
\acrodef{6g}[6G]{the sixth generation}
\acrodef{bs}[BS]{base station}
\acrodef{bse}[BSE]{beam squint effect}
\acrodef{cp}[CP]{cyclic prefix}
\acrodef{elaa}[ELAA]{extremely large antenna array}
\acrodef{ff}[FF]{far-field}
\acrodef{las}[L\&S]{localization and sensing}
\acrodef{los}[LOS]{line-of-sight}
\acrodef{nf}[NF]{near-field}
\acrodef{ris}[RIS]{reconfigurable intelligent surface}
\acrodef{rtt}[RTT]{round-trip-time}
\acrodef{sns}[SNS]{spatial non-stationarity}
\acrodef{swm}[SWM]{spherical wave model}

\acrodef{siso}[SISO]{single-input-single-output}
\acrodef{mimo}[MIMO]{multi-input-multi-output}
\acrodef{ue}[UE]{user equipment}
\acrodef{dmimo}[D-MIMO]{distributed MIMO}
\acrodef{sp}[SP]{scatter point}
\acrodef{awgn}[AWGN]{additive white Gaussian noise}
\acrodef{nlos}[NLOS]{non-line-of-sight}
\acrodef{ofdm}[OFDM]{orthogonal frequency division multiplexing}
\acrodef{tdoa}[TDOA]{time-difference-of-arrival}
\acrodef{toa}[TOA]{time-of-arrival}
\acrodef{am}[AM]{artificial multipath}
\acrodef{an}[AN]{artificial noise}
\acrodef{csi}[CSI]{channel state information}
\acrodef{mcrb}[MCRB]{misspecified Cramér-Rao bound}
\acrodef{crb}[CRB]{Cramér-Rao bound}
\acrodef{lb}[LB]{lower bound}
\acrodef{rmse}[RMSE]{root mean squared error}
\acrodef{psd}[PSD]{power spectral density}
\acrodef{pdf}[PDF]{probability distribution function}
\acrodef{aoa}[AOA]{angle-of-arrival}
\acrodef{aod}[AOD]{angle-of-departure}
\acrodef{fim}[FIM]{Fisher information matrix}
\acrodef{crb}[CRB]{Cramér-Rao bound}
\acrodef{moo}[MOO]{multi-objective optimization}
\acrodef{qos}[QoS]{quality of service}
\acrodef{sdp}[SDP]{semi-definite programming}
\acrodef{lmi}[LMI]{linear matrix inequality}
\acrodef{sdr}[SDR]{semi-definite relaxation}
\acrodef{rcs}[RCS]{radar cross section}
\acrodef{isac}[ISAC]{integrated sensing and communication}
\acrodef{pdd}[PDD]{penalty dual decomposition}
\acrodef{bcd}[BCD]{block coordinate descent}
\usepackage{pgfplots}
\usepackage{tikz}
\usetikzlibrary{calc}
\makeatletter
\newcommand{\gettikzxy}[3]{%
  \tikz@scan@one@point\pgfutil@firstofone#1\relax
  \edef#2{\the\pgf@x}%
  \edef#3{\the\pgf@y}%
}
\newcommand{\quot}[1]{``{#1}''}
\usetikzlibrary{spy,backgrounds}
\usepackage{booktabs} 
\usepackage{amsmath}
\usepackage{graphicx} 
\usepackage{epstopdf}
\usepackage{amssymb}
\usepackage{amsfonts}
\usepackage{amsthm}
\usepackage{cite}
\usepackage{bm,comment}
\usepackage{algorithm}
\usepackage{algpseudocode}

\usepackage{subeqnarray}
\usepackage{subfigure}
\usepackage{multicol}
\usepackage{multirow}
\usepackage{diagbox}
\usepackage{slashbox}
\usepackage{stfloats}
\usepackage{float}
\usepackage{color} 
\usepackage{cases}
\usepackage{lipsum}

\usepackage{geometry}
\allowdisplaybreaks
\begin{document}
\setlength{\textfloatsep}{4pt}
\bstctlcite{IEEEexample:BSTcontrol}
\title{Privacy Preservation in MIMO-OFDM Localization Systems: A Beamforming Approach}
\author{Yuchen Zhang, Hui Chen, Musa Furkan Keskin, Alireza Pourafzal, Pinjun Zheng,  \\
Henk Wymeersch, \emph{Fellow, IEEE}, and 
Tareq Y. Al-Naffouri, \emph{Fellow, IEEE}
\vspace{-0.5cm}
\thanks{
Y. Zhang and T. Y. Al-Naffouri are with the Electrical and Computer Engineering Program, Computer, Electrical and Mathematical Sciences and Engineering (CEMSE), King Abdullah University of Science and Technology (KAUST), Thuwal 23955-6900, Kingdom of Saudi Arabia (e-mail: \{yuchen.zhang; tareq.alnaffouri\}@kaust.edu.sa). H. Chen, M. F. Keskin, A. Pourafzal, and H. Wymeersch are with the Department of Electrical Engineering, Chalmers University of Technology, 41296 Gothenburg, Sweden (e-mail: \{hui.chen; furkan; alireza.pourafzal; henkw\}@chalmers.se). P. Zheng is with the School of Engineering, the University of British Columbia, Kelowna, BC, Canada (e-mail: pinjun.zheng@ubc.ca).}
\thanks{This work was supported in part by the King Abdullah University of Science and Technology (KAUST) Office of Sponsored Research (OSR) under Award ORA-CRG2021-4695, and by the European Commission through the Horizon Europe/JU SNS project Hexa-X-II (Grant Agreement no. 101095759), in part by the Swedish Research Council (VR grant 2023-03821), and in part by the Chalmers Transport Area of Advance project \quot{Towards a Multi-Layer Security Vision for Transportation Systems in the 6G Era}.}
}
\maketitle

\begin{abstract}
We investigate an uplink MIMO-OFDM localization scenario where a legitimate base station (BS) aims to localize a user equipment (UE) using pilot signals transmitted by the UE, while an unauthorized BS attempts to localize the UE by eavesdropping on these pilots, posing a risk to the UE’s location privacy. To enhance legitimate localization performance while protecting the UE’s privacy, we formulate an optimization problem regarding the beamformers at the UE, aiming to minimize the Cramér-Rao bound (CRB) for legitimate localization while constraining the CRB for unauthorized localization above a threshold. A penalty dual decomposition optimization framework is employed to solve the problem, leading to a novel beamforming approach for location privacy preservation. Numerical results confirm the effectiveness of the proposed approach and demonstrate its superiority over existing benchmarks.
\end{abstract}

\IEEEpeerreviewmaketitle
\section{Introduction}
Location information is becoming increasingly vital, enabling a wide range of applications such as digital twins, autonomous driving, and more\cite{henk2024tutorial}. Although global navigation satellite systems have been widely used, they often fall short in environments with poor satellite visibility, leading to the rise of radio localization through 5G/6G cellular networks to provide seamless localization services\cite{henk2022cl,henk2024tutorial}. However, location data can reveal highly sensitive information, such as personal activities, raising significant privacy concerns. The inherent openness of wireless propagation leaks location information at unauthorized nodes, creating privacy threats and highlighting the need for advanced methods to safeguard users' location data in these evolving systems.


Physical layer security, which aims to protect communication from eavesdropping, has been widely studied \cite{physec2017survey}. However, in scenarios where location privacy preservation is crucial, such as in Internet-of-Things localization applications \cite{iotsurvey2022}, the sole focus on communication security may not align with the primary objectives. In \cite{roth2021localization}, it is shown that attackers could exploit the chosen precoder to localize users based on the location of \ac{bs}, suggesting that random selection among precoders that ensure high transmission rates could be an effective countermeasure. Additionally, techniques such as pilot signal modification have been proposed to prevent unauthorized localization \cite{jianxiu2024tsp, zack2024privacy, studer2024twc}. However, this approach might not be practical, as standardized systems require all users to utilize predefined pilots.

\begin{figure}[t]
\centering
\includegraphics[width=0.8\linewidth]{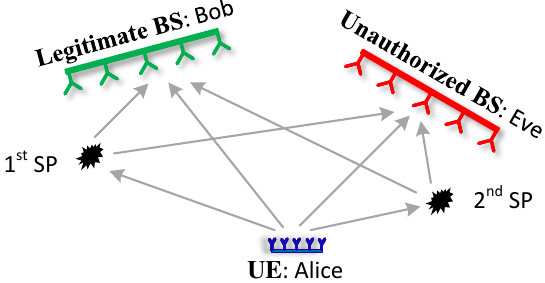}
\caption{Illustration of the risk of location privacy leakage in a MIMO OFDM localization system with coexisting legitimate and unauthorized nodes.}\label{sys_mod}
\end{figure}

Multi-antenna beamforming technologies have been widely adopted to enhance secure communication by flexibly reconfiguring the spatial distribution of signal power. While extensive research has focused on securing communication information through beamforming \cite{zack2024tsp, physec2017survey}, its role in preserving location privacy has received comparatively less attention. In \cite{stefano2020cl} and \cite{stefano2022icc}, beamforming schemes are proposed that enable the \ac{ue} to communicate securely with the \ac{bs} without disclosing its location. 
However, these approaches often {1)} struggle to balance legitimate localization with location privacy protection, and {2)} do not provide a direct metric such as the \ac{crb} for location privacy, relying on metrics like signal-to-noise ratio. 
The challenge of optimizing beamforming to enhance legitimate localization performance while quantitatively safeguarding location privacy has received limited attention.

In this paper, we investigate a \ac{mimo}-\ac{ofdm} uplink localization scenario, where a legitimate \ac{bs} (Bob) aims to localize a \ac{ue} (Alice) based on the received pilot signals. However, these uplink pilots are also intercepted by an unauthorized \ac{bs} (Eve), leading to the risk of Alice’s location information being leaked. The key contributions are summarized as follows: 
\textit{(i)} We formulate an optimization problem to provide a reliable localization of Alice for Bob, while quantitatively protecting Alice's location privacy from Eve. The problem minimizes the \ac{crb} of legitimate localization, subject to constraining the \ac{crb} of the unauthorized localization above a predefined threshold.
\textit{(ii)} We address the non-convex problem using matrix lifting and the \ac{pdd} optimization framework, introducing a novel beamforming technique that enhances legitimate localization performance while maintaining designated privacy levels.
\textit{(iii)} We demonstrate with numerical results the superior performance of the proposed beamforming approach compared to other benchmarks.

\section{System Model And Problem Formulation} 
As shown in Fig. \ref{sys_mod}, we consider a \ac{mimo}-\ac{ofdm}-based uplink localization system with $M$ subcarriers. Alice equipped with $M_{\text{A}}$ transmit antennas, sends publicly known \ac{ofdm} pilot signals over $L$ time slots. Bob with $M_{\text{B}}$ antennas, processes the received signals to estimate Alice's location. However, due to the inherent broadcast nature of wireless communication, Eve equipped with $M_{\text{E}}$ antennas, can also intercept the signals and perform the same localization, leading to unwanted leakage of Alice’s location information. 

We assume that the locations of both Bob and Eve are known to Alice, as they typically have fixed locations as \acp{bs}. Additionally, the channels from Alice to Bob and from Alice to Eve are both influenced by the same group of \acp{sp}. For the sake of conciseness, we use Bob's localization as an example and derive the performance based on the \ac{crb}. Due to the symmetry of the problem, the localization performance at Eve can be obtained directly. 

\subsection{Signal Model}
Let $N$ represent the number of \ac{ofdm} pilot symbols per slot. The transmitted signal for the $n$-th symbol in the $l$-th slot over the $m$-th subcarrier is expressed as 
\begin{equation}\label{tran-sig}
\boldsymbol{x}\left[l,n,m\right] = \boldsymbol{w}\left[l\right]s\left[n,m\right],
\end{equation}
where $\boldsymbol{w}\left[l\right]\in \mathbb{C}^{M_{\text{B}}}$ is the beamformer in the $l$-th slot, and $s\left[n,m\right]$ is the unit-modulus pilot symbol on subcarrier $m$. The Alice-to-Bob channel for subcarrier $m$ is modeled as
\begin{equation}\label{bob-chan}
\boldsymbol{H}_{\text{B}}\left[m\right] = \sum_{k = 0}^{K} \alpha_{\text{B}, k}e^{-\jmath 2 \pi m \Delta f \tau_{\text{B},k}} \boldsymbol{a}_{\text{B}}\left(\theta_{\text{B},k}\right)
\boldsymbol{a}_{\text{A}}^{\mathsf{H}}\left(\theta_{\text{A},k}\right),
\end{equation}
where $K$ denotes the number of \acp{sp}, while $\alpha_{\text{B},k}$, $\tau_{\text{B},k}$, $\theta_{\text{B},k}$, and $\theta_{\text{A},k}$ refer to the complex gain, delay, \ac{aoa}, and \ac{aod} of the $k$-th \ac{sp}, respectively. Note that for simplicity, the \ac{los} path is indexed by $k=0$. Here, $\theta_{\text{B},0}$ and $\theta_{\text{A},0}$ represent the \ac{aoa} and \ac{aod} at Bob and Alice, respectively. Finally, $\boldsymbol{a}_{\text{A}}\left(\theta\right) \in \mathbb{C}^{M_{\text{A}}}$ and $\boldsymbol{a}_{\text{B}}\left(\theta\right) \in \mathbb{C}^{M_{\text{B}}}$ denote the steering vectors at Alice and Bob, respectively. The received signal at Bob is
\begin{equation}\label{re-sig-bob}
\boldsymbol{y}_{\text{B}}\left[l,n,m\right] = \boldsymbol{H}_{\text{B}}\left[m\right]\boldsymbol{x}\left[l,n,m\right] +\boldsymbol{z}_{\text{B}}\left[l,n,m\right],
\end{equation}
where $\boldsymbol{z}_{\text{B}}\left[l,n,m\right] \sim  \mathcal{CN}(\boldsymbol{0},\sigma^2 \boldsymbol{F}_{M_{\text{B}}})$ is the \ac{awgn} at Bob's receiver. Here, $\sigma^2 = F N_0 \Delta f$ is the noise power, where $F$, $N_0$, and $\Delta f$ represent the noise figure, single-sided \ac{psd}, and subcarrier spacing, respectively.

\subsection{CRB-Based Performance Metric}

The channel domain parameters relevant to Alice-Bob link are represented as $\boldsymbol{\xi}_{\text{B}} = [\boldsymbol{\theta}_{\text{A}}^{\mathsf{T}},\boldsymbol{\theta}_{\text{B}}^{\mathsf{T}},\boldsymbol{\tau}_{\text{B}}^{\mathsf{T}},\boldsymbol{\alpha}_{\text{B},\text{R}}^{\mathsf{T}},\boldsymbol{\alpha}_{\text{B},\text{I}}^{\mathsf{T}}]^{\mathsf{T}} \in \mathbb{R}^{(5K+5)}$. Here, $\boldsymbol{\theta}_{\text{A}} = [{\theta}_{\text{A}, 0}, \ldots, {\theta}_{\text{A}, K}] \in \mathbb{R}^{(K+1)}$ denotes the angles of departure (AODs), while $\boldsymbol{\theta}_{\text{B}} = [{\theta}_{\text{B}, 0}, \ldots, {\theta}_{\text{B}, K}] \in \mathbb{R}^{(K+1)}$ refers to the angles of arrival (AOAs). The delay measurements are encapsulated in $\boldsymbol{\tau}_{\text{B}} = [{\tau}_{\text{B},0}, \ldots, {\tau}_{\text{B},K}] \in \mathbb{R}^{(K+1)}$, and the real and imaginary parts of the complex channel gains are captured in $\boldsymbol{\alpha}_{\text{B},\text{R}} = [\Re\{{\alpha}_{0}\}, \ldots, \Re\{{\alpha}_{K}\}] \in \mathbb{R}^{(K+1)}$ and $\boldsymbol{\alpha}_{\text{B},\text{I}} = [\Im\{{\alpha}_{\text{B},0}\}, \ldots, \Im\{{\alpha}_{\text{B},K}\}] \in \mathbb{R}^{(K+1)}$, respectively. Utilizing the Slepian-Bangs formula\cite{furkan2022tvt}, the $(i,j)$-th element of the channel-domain \ac{fim} $\boldsymbol{F}_{\text{c}}(\boldsymbol{\xi}_{\text{B}})$ can be expressed as
\begin{align}
&\left[\boldsymbol{F}_{\text{c}}\left(\boldsymbol{\xi}_{\text{B}}\right)\right]_{i,j}
\!=\! \frac{2}{\sigma^2}\!\sum_{l=1}^{L}\!\sum_{n=1}^{N}\!
\sum_{m=1}^{M}\!\Re\! \left\{\!\frac{\partial \boldsymbol{\mu}_{\text{B}}\left[l,n,m\right]^{\mathsf{H}}}{\partial \left[\boldsymbol{\xi}_{\text{B}}\right]_{i}}\frac{\partial \boldsymbol{\mu}_{\text{B}}\left[l,n,m\right]}{\partial \left[\boldsymbol{\xi}_{\text{B}}\right]_{j}}\!\right\}\notag\\
&= \frac{2N}{\sigma^2}\sum_{m=1}^{M}\Re \left\{\text{tr}\left(\frac{\partial \boldsymbol{H}_{\text{B}}\left[m\right]}{\partial \left[\boldsymbol{\xi}_{\text{B}}\right]_{j}} \boldsymbol{W}\boldsymbol{W}^{\mathsf{H}}\frac{\partial \boldsymbol{H}_{\text{B}}\left[m\right]^{\mathsf{H}}}{\partial \left[\boldsymbol{\xi}_{\text{B}}\right]_{i}}\right)\right\},\label{bob-chan-fim}
\end{align}
where $\boldsymbol{\mu}_{\text{B}}\left[l,n,m\right] = \boldsymbol{H}_{\text{B}}\left[m\right]\boldsymbol{x}\left[l,n,m\right]$ denotes the noise-free observation from \eqref{re-sig-bob}, and $\boldsymbol{W} = [\boldsymbol{w}_1,\ldots,\boldsymbol{w}_L] \in \mathbb{C}^{M_{\text{B}} \times L}$ encompasses $L$ beamformers.

Since we focus on the localization performance, the location-domain parameters are consolidated as $\boldsymbol{\eta}_{\text{B}} = [\boldsymbol{p}_{\text{A}}^{\mathsf{T}},\phi_{\text{B}},\boldsymbol{p}_{1}^{\mathsf{T}},\ldots,\boldsymbol{p}_{K}^{\mathsf{T}},\Delta t_{\text{B}}, \boldsymbol{\alpha}_{\text{B},\text{R}}^{\mathsf{T}},\boldsymbol{\alpha}_{\text{B},\text{I}}^{\mathsf{T}}] \in \mathbb{R}^{(4K+6)}$, where $\boldsymbol{p}_{\text{A}} \in \mathbb{R}^{2}$ indicates Alice's position, and $\boldsymbol{p}_{k} \in \mathbb{R}^{2}$ denotes the location of the $k$-th scatterer. The variable $\phi_{\text{B}}$ represents Alice's relative orientation in Bob's local coordinate system, while $\Delta t_{\text{B}}$ accounts for the clock bias, reflecting the timing mismatch between Alice and Bob. Notably, the nuisance parameters $\boldsymbol{\alpha}_{\text{B},\text{R}}$ and $\boldsymbol{\alpha}_{\text{B},\text{I}}$ derived from the channel-domain parameter $\boldsymbol{\xi}_{\text{B}}$ persist in the location-domain parameter $\boldsymbol{\eta}_{\text{B}}$, as they do not provide beneficial information for estimating position. The location-domain FIM, $\boldsymbol{F}_{\text{p}}(\boldsymbol{\eta}_{\text{B}})$, is calculated from the channel-domain FIM as follows
\begin{equation}\label{bob-pos-fim}
\boldsymbol{F}_{\text{p}}\left(\boldsymbol{\eta}_{\text{B}}\right) = \boldsymbol{J}_{\text{B}}^{\mathsf{T}} \boldsymbol{F}_{\text{c}}\left(\boldsymbol{\xi}_{\text{B}}\right)   \boldsymbol{J}_{\text{B}},
\end{equation}
where $\boldsymbol{J}_{\text{B}} \in  \mathbb{R}^{(5K+5)\times (4K+6)}$ represents the Jacobian matrix, with its $(i,j)$-th element defined as $[\boldsymbol{J}_{\text{B}}]_{i,j} = \partial [\boldsymbol{\xi}_{\text{B}}]_{i} / \partial [\boldsymbol{\eta}_{\text{B}}]_{j}$. The \ac{crb} is utilized to assess the localization precision w.r.t. $\boldsymbol{p}_{\text{A}}$ at Bob, offering a lower limit on the total variances for estimating $\boldsymbol{p}_{\text{A}}$, expressed as follows
\begin{equation}\label{crb-pos-bob}
\text{CRB}_{\text{B}}\left(\boldsymbol{p}_{\text{A}}\right) = \text{tr}\left(\left[\boldsymbol{F}_{\text{p}}\left(\boldsymbol{\eta}_{\text{B}}\right)^{-1}\right]_{1 : 2, 1 : 2}\right).    
\end{equation}

\subsection{Problem Formulation}
From \eqref{bob-chan-fim}, it is clear that the \ac{crb} for localization performance at Bob, i.e., $\text{CRB}_{\text{B}}(\boldsymbol{p}_{\text{A}})$, depends on the design of $\boldsymbol{W}$, which can be optimized through appropriate beamformer configurations. At the same time, the \ac{crb} for localization performance at Eve, represented as $\text{CRB}_{\text{E}}(\boldsymbol{p}_{\text{A}})$, also depends on $\boldsymbol{W}$. To preserve location privacy, it is essential to enhance legitimate localization performance while limiting the performance at the unauthorized node. Hence, we formulate an optimization problem about $\boldsymbol{W}$ that seeks to minimize the legitimate \ac{crb}, ensuring that the \ac{crb} at the unauthorized node remains above a specified threshold\footnote{Note that the current optimization framework requires knowledge of parameters associated with \ac{crb} calculation, such as the UE's and \acp{sp}' positions, clock bias, and orientations, obtained from external sensors or tracking mechanisms. However, practical scenarios may involve inaccurate estimation of these parameters. Such cases can be addressed within a similar framework to \eqref{opt-crb} using robust methods, as in \cite{furkan2022tvt,henk2022jstsp}. This scenario, however, is beyond the scope of this paper and left for future study.}, expressed as
\begin{subequations}\label{opt-crb}
\begin{align}
\mathop {\min }\limits_{\boldsymbol{W}} \;\;\; &\text{CRB}_{\text{B}}\left(\boldsymbol{p}_{\text{A}}\right)\label{opt-crb-obj}\\
{\rm{s.t.}}\;\;\;
& \text{CRB}_{\text{E}}\left(\boldsymbol{p}_{\text{A}}\right) \ge \gamma,\label{opt-crb-privacy}
\\
&  \text{tr}\left(\boldsymbol{W}\boldsymbol{W}^{\mathsf{H}}\right) \le P/M, \label{opt-crb-power}
\end{align}
\end{subequations}
where $\gamma$ is Eve's \ac{crb} threshold, determined by the practical requirement, and $P$ denotes the power budget. We set the right side of \eqref{opt-crb-power} to $P/M$, ensuring that the total transmit power across subcarriers equals $P$. Note that both $\text{CRB}_{\text{B}}(\boldsymbol{p}_{\text{A}})$ and $\text{CRB}_{\text{E}}(\boldsymbol{p}_{\text{A}})$ are non-convex and non-concave functions of $\boldsymbol{W}$, respectively, complicating solving \eqref{opt-crb}. 

\section{PDD-Based Beamforming for location privacy Preservation}
In the following, we employ matrix lifting and propose a scheme based on the \ac{pdd} optimization framework \cite{qingjiang2020tsp}, solving \eqref{opt-crb} iteratively. 

\subsection{Problem Reformulation}
Note that the matrices on the right-hand side of \eqref{crb-pos-bob} can be reformulated as \cite{henk2019twc}
\begin{equation}\label{efim}
\left[\boldsymbol{F}_{\text{p}}\left(\boldsymbol{\eta}_{\text{B}}\right)^{-1}\right]_{1 : 2, 1 : 2} = \left[\boldsymbol{Q} - \boldsymbol{G}\boldsymbol{Z}^{-1}\boldsymbol{G}^{\mathsf{T}}\right]^{-1}    
\end{equation}
where $\boldsymbol{Q} = [\boldsymbol{F}_{\text{p}}(\boldsymbol{\eta}_{\text{B}})]_{1 : 2, 1 : 2}$, $\boldsymbol{G} = [\boldsymbol{F}_{\text{p}}(\boldsymbol{\eta}_{\text{B}})]_{1 : 2, 3 : 4K + 6}$, and $\boldsymbol{Z} = [\boldsymbol{F}_{\text{p}}(\boldsymbol{\eta}_{\text{B}})]_{3 : 4K + 6, 3 : 4K + 6}$. Let $\boldsymbol{V} = \boldsymbol{W}\boldsymbol{W}^{\mathsf{H}}$. The elements within $\boldsymbol{Q}$, $\boldsymbol{G}$, and $\boldsymbol{Z}$ become linear with respect to $\boldsymbol{V}$, as inferred from \eqref{bob-chan-fim} and \eqref{bob-pos-fim}. By introducing the auxiliary variable $\boldsymbol{U} \in \mathbb{R}^{2 \times 2}$, we lift \eqref{opt-crb} into an equivalent form as
\begin{subequations}\label{opt-crb-refor}
\begin{align}
\mathop {\min }\limits_{\boldsymbol{V},\boldsymbol{U}} \;\;\; & \text{tr}\left(\boldsymbol{U}^{-1}\right) 
\label{opt-crb-refor-obj}\\
{\rm{s.t.}}\;\;\;
& 
\begin{bmatrix}
\boldsymbol{Q} - \boldsymbol{U} & \boldsymbol{G}\\
\boldsymbol{G}^{\mathsf{T}} &\boldsymbol{Z}
\end{bmatrix} \succeq \mathbf{0}, \quad \boldsymbol{U} \succeq \mathbf{0}, \label{opt-crb-refor-lmi}\\
& \text{CRB}_{\text{E}}\left(\boldsymbol{p}_{\text{A}}\right) \ge \gamma,\label{opt-crb-refor-privacy}
\\
&  \text{tr}\left(\boldsymbol{V}\right) \le P/M,\\
& \text{rank}\left(\boldsymbol{V}\right) = L. \label{opt-crb-refor-rank}
\end{align}
\end{subequations}
Due to the existence of non-convex constraints \eqref{opt-crb-refor-privacy} and \eqref{opt-crb-refor-rank}, \eqref{opt-crb-refor} remains challenging to solve.   

\subsection{PDD-Based Optimization Framework} 
By recalling that $\text{CRB}_{\text{E}}(\boldsymbol{p}_{\text{A}}) = \text{tr}([\boldsymbol{F}_{\text{p}}(\boldsymbol{\eta}_{\text{E}})^{-1}]_{1 : 2, 1 : 2})$ and introducing the auxiliary variable $\boldsymbol{\Phi} \in \mathbb{R}^{(5K+5) \times (5K+5)}$, while ignoring \eqref{opt-crb-refor-rank}, we can relax \eqref{opt-crb-refor} into
\begin{subequations}\label{opt-crb-relax}
\begin{align}
\mathop {\min }\limits_{\boldsymbol{V},\boldsymbol{U},\boldsymbol{\Phi}} \;\;\; & \text{tr}\left(\boldsymbol{U}^{-1}\right) 
\label{opt-crb-relax-obj}\\
{\rm{s.t.}}\;\;\;
& 
\begin{bmatrix}
\boldsymbol{Q} - \boldsymbol{U} & \boldsymbol{G}\\
\boldsymbol{G}^{\mathsf{T}} &\boldsymbol{Z}
\end{bmatrix} \succeq \mathbf{0}, \quad \boldsymbol{U} \succeq \mathbf{0}, \label{opt-crb-relax-lmi}\\
& \boldsymbol{\Phi}_{1,1} + \boldsymbol{\Phi}_{2,2} \ge \gamma, \quad \boldsymbol{\Phi} \succeq \mathbf{0},\label{opt-crb-relax-privacy}
\\
& \boldsymbol{F}_{\text{p}}\left(\boldsymbol{\eta}_{\text{E}}\right) \boldsymbol{\Phi} = \boldsymbol{I},\label{opt-crb-relax-eq}
\\
&  \text{tr}\left(\boldsymbol{V}\right) \le P/M,
\end{align}
\end{subequations}
where $\boldsymbol{F}_{\text{p}}(\boldsymbol{\eta}_{\text{E}}) \in \mathbb{R}^{(5K+5) \times (5K+5)}$ is the corresponding location-domain \ac{fim} from Alice to Eve, whose elements are also linear with $\boldsymbol{V}$, and $\boldsymbol{I}$ denotes the identity matrix such that $\boldsymbol{\Phi}$ serves as the inverse matrix of $\boldsymbol{F}_{\text{p}}(\boldsymbol{\eta}_{\text{E}})$. We note that the challenge in solving \eqref{opt-crb-relax} lies in the non-convex equality constraint \eqref{opt-crb-relax-eq}, which can be effectively addressed using \ac{pdd}. According to \cite{qingjiang2020tsp}, the standard \ac{pdd} optimization framework is developed in a double-loop structure, where the augmented Lagrangian problem (ALP) of the original problem is optimized in a \ac{bcd} manner in the inner loop, while the Lagrangian dual variables and penalty factors are updated in the outer loop. 


\renewcommand{\algorithmicrequire}{\textbf{Input:}}
\renewcommand{\algorithmicensure}{\textbf{Output:}}
\begin{algorithm}[t]
\caption{PDD-Based Algorithm for Solving \eqref{opt-crb-relax}}
\label{pdd-overall}       %
\begin{algorithmic}[1]
\State \textbf{Initialize}: $\boldsymbol{\Theta}=\boldsymbol{0}$, $\rho$, $\delta$, $k = 1$;
\Repeat
\State {Optimize $\left(\boldsymbol{V},\boldsymbol{U},\boldsymbol{\Phi}\right)$ via \ac{bcd};}
\If {$h\left(\boldsymbol{V}, \boldsymbol{\Phi}\right) \le \zeta\left[k\right] $}
\State {$\boldsymbol{\Phi} = \boldsymbol{\Phi} + \rho \left( \boldsymbol{F}_{\text{p}}\left(\boldsymbol{\eta}_{\text{E}}\right) \boldsymbol{\Phi} - \boldsymbol{I} \right)$;}
\Else 
\State {$\rho = \delta \rho$;}
\EndIf
\State {$k = k + 1$}
\Until {$h\left(\boldsymbol{V}, \boldsymbol{\Phi}\right)$ is below a specified threshold.}
\State \textbf{Output}: {$\boldsymbol{V}$, $\boldsymbol{U}$, $\boldsymbol{\Phi}$.} 
\end{algorithmic}
\end{algorithm}

\subsubsection{Augmented Lagrangian Problem}
For the inner loop of the \ac{pdd} framework, the ALP for \eqref{opt-crb-relax} is given by
\begin{subequations}\label{opt-crb-al}
\begin{align}
\mathop {\min }\limits_{\boldsymbol{V},\boldsymbol{U},\boldsymbol{\Phi}} \;\;\; & \text{tr}\left(\boldsymbol{U}^{-1}\right) + \frac{\rho}{2} \left\|  \boldsymbol{F}_{\text{p}}\left(\boldsymbol{\eta}_{\text{E}}\right) \boldsymbol{\Phi} - \boldsymbol{I} + \frac{1}{\rho} \boldsymbol{\Theta}\right\|_{\text{F}}^2 
\label{opt-crb-al-obj}\\
{\rm{s.t.}}\;\;\;
& 
\begin{bmatrix}
\boldsymbol{Q} - \boldsymbol{U} & \boldsymbol{G}\\
\boldsymbol{G}^{\mathsf{T}} &\boldsymbol{Z}
\end{bmatrix} \succeq \mathbf{0}, \quad \boldsymbol{U} \succeq \mathbf{0}, \label{opt-crb-al-lmi}\\
& \boldsymbol{\Phi}_{1,1} + \boldsymbol{\Phi}_{2,2} \ge \gamma, \quad \boldsymbol{\Phi} \succeq \mathbf{0},\label{opt-crb-al-privacy}
\\
&  \text{tr}\left(\boldsymbol{V}\right) \le P/M,
\end{align}
\end{subequations}
where $\boldsymbol{\Theta} \in \mathbb{R}^{(5K+5) \times (5K+5)}$ and $\rho$ are the Lagrangian dual variable and penalty factor, respectively. 

\subsubsection{Solving ALP with BCD}
It can be observed that, with fixed $\boldsymbol{\Theta}$ and $\rho$, \eqref{opt-crb-al} remains non-convex due to the presence of the bilinear term $\boldsymbol{F}_{\text{p}}(\boldsymbol{\eta}_{\text{E}}) \boldsymbol{\Phi}$, as $\boldsymbol{F}_{\text{p}}(\boldsymbol{\eta}_{\text{E}})$ is linear with $\boldsymbol{V}$. However, by leveraging the technique of \ac{bcd}, the augmented Lagrangian problem \eqref{opt-crb-al} can be solved alternately. In particular, with fixed $\boldsymbol{\Phi}$, the convex \ac{sdp} subproblem w.r.t. $(\boldsymbol{V},\boldsymbol{U})$ is given by
\begin{subequations}\label{opt-crb-al-vu}
\begin{align}
\mathop {\min }\limits_{\boldsymbol{V},\boldsymbol{U}} \;\;\; & \text{tr}\left(\boldsymbol{U}^{-1}\right) + \frac{\rho}{2} \left\|  \boldsymbol{F}_{\text{p}}\left(\boldsymbol{\eta}_{\text{E}}\right) \boldsymbol{\Phi} - \boldsymbol{I} + \frac{1}{\rho} \boldsymbol{\Theta}\right\|_{\text{F}}^2 \\
{\rm{s.t.}}\;\;\;
& 
\begin{bmatrix}
\boldsymbol{Q} - \boldsymbol{U} & \boldsymbol{G}\\
\boldsymbol{G}^{\mathsf{T}} &\boldsymbol{Z}
\end{bmatrix} \succeq \mathbf{0}, \quad \boldsymbol{U} \succeq \mathbf{0}, \\
&  \text{tr}\left(\boldsymbol{V}\right) \le P/M.
\end{align}
\end{subequations}
Next, with fixed $(\boldsymbol{V},\boldsymbol{U})$, the convex \ac{sdp} subproblem w.r.t. $\boldsymbol{\Phi}$ is given by
\begin{subequations}\label{opt-crb-al-phi}
\begin{align}
\mathop {\min }\limits_{\boldsymbol{\Phi}} \;\;\; & \left\|  \boldsymbol{F}_{\text{p}}\left(\boldsymbol{\eta}_{\text{E}}\right) \boldsymbol{\Phi} - \boldsymbol{I} + \frac{1}{\rho} \boldsymbol{\Theta}\right\|_{\text{F}}^2 
\label{opt-crb-al-phi-obj}\\
{\rm{s.t.}}\;\;\;
& \boldsymbol{\Phi}_{1,1} + \boldsymbol{\Phi}_{2,2} \ge \gamma, \quad \boldsymbol{\Phi} \succeq \mathbf{0}.
\end{align}
\end{subequations}
Following the principles of \ac{bcd}, \eqref{opt-crb-al} can be addressed by alternately solving \eqref{opt-crb-al-vu} and \eqref{opt-crb-al-phi} in an iterative manner until convergence.

We define the violation function as $h(\boldsymbol{V},\boldsymbol{\Phi}) = \left\|\boldsymbol{F}_{\text{p}}\left(\boldsymbol{\eta}_{\text{E}}\right) \boldsymbol{\Phi} - \boldsymbol{I}\right\|_{\infty}$. The steps of \eqref{opt-crb-relax} using the \ac{pdd} approach is summarized in Algorithm \ref{pdd-overall}. Here, $\delta > 1$ is a constant that increases the penalty factor when necessary, while $\zeta[k]$ denotes a sequence determined empirically to approach zero. Specifically, we define $\zeta[k] = q h^{(k-1)}(\boldsymbol{V},\boldsymbol{\Phi})$, where $q \in (0,1)$ is an attenuation constant, and $h^{(k-1)}(\boldsymbol{V},\boldsymbol{\Phi})$ is the value of $h(\boldsymbol{V},\boldsymbol{\Phi})$ at the $(k-1)$-th iteration. 

After obtaining $\boldsymbol{V}$ from Algorithm \ref{pdd-overall}, the beamformers $\boldsymbol{W}$ can then be derived from $\boldsymbol{V}$ via matrix decomposition or randomization techniques \cite{tom2010spm}.

\subsection{Convergence and Complexity}
As $\rho$ increases, the term $\left\|  \boldsymbol{F}_{\text{p}}\left(\boldsymbol{\eta}_{\text{E}}\right) \boldsymbol{\Phi} - \boldsymbol{I} + \frac{1}{\rho} \boldsymbol{\Theta}\right\|_{\text{F}}$ tends to zero, indicating that the constraint \eqref{opt-crb-relax-eq} in \eqref{opt-crb-relax} is satisfied over iterations. A detailed discussion on the convergence of the \ac{pdd} framework is provided in \cite{qingjiang2020tsp}, which ensures the convergence of Algorithm \ref{pdd-overall}. 

The overall complexity of Algorithm \ref{pdd-overall} is primarily dictated by the inner \ac{bcd} process. According to \cite{furkan2022tvt}, the complexity of an \ac{sdp} problem is $\mathcal{O}(I^2 \sum_{j=1}^{J}d_j^2 + I \sum_{j=1}^{J}d_j^3)$, where $I$ and $J$ denote the number of variables and \ac{lmi} constraints, and $d_j$ represents the size of the $j$-th matrix. For subproblem \eqref{opt-crb-al-vu}, the complexity is approximated as $\mathcal{O}(M_{\text{A}}^4 K^2 + M_{\text{A}}^2 K^3)$, and for subproblem \eqref{opt-crb-al-phi}, it is $\mathcal{O}(K^6)$.

\begin{figure*}[t]
\centering
\centerline{\includegraphics[width=1\linewidth]{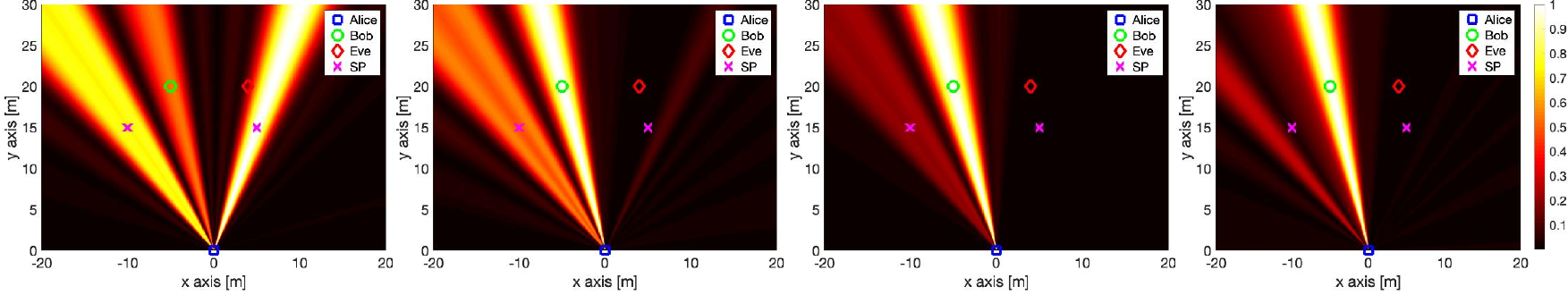}}
\small
\centerline{(a) $\sqrt{\gamma} = 0\text{ m}$ \hspace{2.4cm} (b) $\sqrt{\gamma} = 1\text{ m}$ \hspace{2.4cm} (c) $\sqrt{\gamma} = 4\text{ m}$ \hspace{2.4cm} (d) $\sqrt{\gamma} = 10\text{ m}$}
\normalsize
\caption{Beampatterns under different location privacy constraint: (a) $\sqrt{\gamma} = 0\text{ m}$; (b) $\sqrt{\gamma} = 1\text{ m}$; (c) $\sqrt{\gamma} = 4\text{ m}$; (d) $\sqrt{\gamma} = 10\text{ m}$.}
\label{bp}
\vspace{-0.5cm}
\end{figure*}

\section{Numerical Results}
\subsection{Scenarios}
The simulation setup assumes the following parameters unless otherwise specified: Alice is equipped with $M_{\text{A}} = 16$ transmit antennas and is located at $\boldsymbol{p}_{\text{A}} = [0\text{ m}, 0\text{ m}]^{\mathsf{T}}$. Bob, also using $M_{\text{B}} = 16$ antennas, is placed at $\boldsymbol{p}_{\text{B}} = [-5\text{ m}, 20\text{ m}]^{\mathsf{T}}$, while Eve, equipped with $M_{\text{E}} = 16$ antennas, is localized at $\boldsymbol{p}_{\text{E}} = [4\text{ m}, 20\text{ m}]^{\mathsf{T}}$. Two \acp{sp} are present at $\boldsymbol{p}_1 = [-10\text{ m}, 15\text{ m}]^{\mathsf{T}}$ and $\boldsymbol{p}_2 = [5\text{ m}, 15\text{ m}]^{\mathsf{T}}$. 
The system's transmit power is $P = -20\text{ dBm}$, with a carrier frequency of $f_c = 28 \text{ GHz}$ and a bandwidth of $W = 120 \text{ MHz}$. The number of subcarriers is $M = 1024$, the noise figure is $F = 10 \text{ dB}$, and the noise power spectral density is $N_0 = -173.855 \text{ dBm/Hz}$. The simulation covers $L = 16$ time slots, each containing $N = 100$ pilot \ac{ofdm} signals. Bob's clock bias is $\Delta t_{\text{B}} = 1 \;\mu\text{s}$, and his relative orientation is $\phi_{\text{B}} = (110/180)\pi$, while Eve's clock bias and orientation are $\Delta t_{\text{E}} = 1 \;\mu\text{s}$ and $\phi_{\text{E}} = (200/180)\pi$, respectively. 
Channel gains follow a free-space path loss model\cite{henk2022cl}. For the Alice-Bob link, the \ac{los} channel gain is $\alpha_{\text{B},0} = e^{\jmath \omega_{\text{B},0}} \lambda/(4 \pi \left\|\boldsymbol{p}_{\text{A}} - \boldsymbol{p}_{\text{B}}\right\|)$, and the \ac{nlos} gain is $\alpha_{\text{B},k} = \sigma_{\text{RCS}} e^{\jmath \omega_{\text{B},k}} \lambda/((4 \pi)^{3/2} \left\|\boldsymbol{p}_{\text{A}} - \boldsymbol{p}_k\right\| \left\|\boldsymbol{p}_k - \boldsymbol{p}_{\text{B}}\right\|)$. Similarly, the \ac{los} gain for the Alice-Eve link is $\alpha_{\text{E},0} = e^{\jmath \omega_{\text{E},0}} \lambda/(4 \pi \left\|\boldsymbol{p}_{\text{A}} - \boldsymbol{p}_{\text{E}}\right\|)$, and the \ac{nlos} gain is $\alpha_{\text{E},k} = \sigma_{\text{RCS}} e^{\jmath \omega_{\text{E},k}} \lambda/((4 \pi)^{3/2} \left\|\boldsymbol{p}_{\text{A}} - \boldsymbol{p}_k\right\| \left\|\boldsymbol{p}_k - \boldsymbol{p}_{\text{E}}\right\|)$. Here, $\omega_{\text{B},k}$ and $\omega_{\text{E},k}$ are the uniformly distributed random phases of Alice-Bob and Alice-Eve links, respectively. The \ac{rcs} for each \ac{sp} is $\sigma_{\text{RCS}} = 100 \text{ m}^2$, and the wavelength is $\lambda = c/f_c$, where $c$ represents the speed of light.
\subsection{Benchmarks}
For comparison, we introduce two power-adjustment-based benchmarks, which are detailed below:
\begin{itemize}
    \item \emph{Benchmark I:} By ignoring the location privacy constraint \eqref{opt-crb-privacy}, \eqref{opt-crb} reduces to a location-domain \ac{crb} minimization problem, which can be solved using the approach proposed in \cite{furkan2022tvt}. If the solution already satisfies \eqref{opt-crb-privacy}, we retain the obtained beamformers. Otherwise, we reduce the transmit power until \eqref{opt-crb-privacy} holds with equality.
    \item \emph{Benchmark II:} We first obtain beamformers using the low-complexity, codebook-based power allocation scheme from \cite{furkan2022tvt,henk2022jstsp}, which exploits the structure of the optimal variance matrix to minimize the \ac{crb}. As in Benchmark I, we then adjust the transmit power based on the same principle.
\end{itemize}
\subsection{Results and Discussion}

\subsubsection{Beampatterns}
Figures \ref{bp}(a)-(d) illustrate the beampatterns (normalized receive powers) of the proposed beamforming approach for different values of $\gamma$, representing varying levels of location privacy requirements. When $\sqrt{\gamma} = 0\text{ m}$, meaning no location privacy constraint is applied, the problem reduces to the \ac{crb}-minimizing problem addressed in \cite{furkan2022tvt}. As shown in Fig. \ref{bp}(a), Bob, serving as the sole anchor with a known position, and two \acp{sp}, which create resolvable paths advantageous for single-anchor localization, are simultaneously illuminated by three beams. As $\gamma$ increases, implying a stricter location privacy constraint, the beam illuminating the \ac{sp} on the right, which is closer to Eve, diminishes. This occurs because energy leakage to Eve must be minimized, thereby protecting Alice's location privacy at the cost of Bob's localization performance. As $\gamma$ increases further, such as when $\sqrt{\gamma}$ reaches $10\text{ m}$, even the beam on the left slightly shifts away from the illuminated \ac{sp}, which could otherwise enhance Eve's localization performance.

\begin{figure}[htb]
\centering
\begin{minipage}[b]{0.98\linewidth}
\vspace{-0.3cm}
  \centering
%
%
\definecolor{mycolor1}{rgb}{0.00000,0.44700,0.74100}%
\definecolor{mycolor2}{rgb}{0.85000,0.32500,0.09800}%
\definecolor{mycolor3}{rgb}{0.92900,0.69400,0.12500}%
\begin{tikzpicture}[font=\footnotesize, spy using outlines={rectangle, magnification=3, size=1cm, connect spies}]
\begin{axis}[%
width=72mm,
height=38mm,
at={(0in,0in)},
scale only axis,
xmin=0,
xmax=10,
xlabel style={font=\color{white!15!black},font=\footnotesize},
xlabel={$\sqrt{\gamma}$ [m]},
ymode=log,
ymin=0,
ymax=1.8,
ylabel style={font=\color{white!15!black},font=\footnotesize},
ylabel={$\sqrt{\text{CRB}}$ at Bob [m]},
axis background/.style={fill=white},
xmajorgrids,
ymajorgrids,
legend columns=3, 
legend style={at={(0.02,1.02)}, anchor=south west, legend cell align=left, align=left, draw=white!15!black}
]
\addplot [color=mycolor1, line width=1.0pt, mark=diamond, mark options={solid, mycolor1}]
  table[row sep=crcr]{%
0	0.0312195616908725\\
0.1	0.0312200674137014\\
0.4	0.0469670701764172\\
0.7	0.0577412984482\\
1	0.0623146491858398\\
2	0.0665385994760365\\
4	0.0754772794553757\\
10	0.126275648885389\\
};
\addlegendentry{Proposed}

\addplot [color=mycolor2, dashed, line width=1.0pt, mark=x, mark options={solid, mycolor2}]
  table[row sep=crcr]{%
0	0.0312195616908725\\
0.1	0.0312195617874693\\
0.4	0.0547116080867174\\
0.7	0.0957453141517554\\
1	0.136779020216793\\
2	0.273558040433587\\
4	0.547116080867174\\
10	1.36779020216793\\
};
\addlegendentry{Benchmark I}

\addplot [color=mycolor3, dashed, line width=1.0pt, mark=x, mark options={solid, mycolor3}]
  table[row sep=crcr]{%
0	0.0587550107189978\\
0.1	0.061110081345082\\
0.4	0.071952974276995\\
0.7	0.125917704984741\\
1	0.179882435692488\\
2	0.359764871384975\\
4	0.71952974276995\\
10	1.79882435692488\\
};
\addlegendentry{Benchmark II}


\end{axis}
\end{tikzpicture}%
    \vspace{-1.cm}
  \centerline{(a) $\sqrt{\text{CRB}}$ at Bob versus $\sqrt{\gamma}$} \medskip
\end{minipage}
\hfill
\begin{minipage}[b]{0.98\linewidth}
\vspace{-0.3cm}
  \centering
%
%
\definecolor{mycolor1}{rgb}{0.00000,0.44700,0.74100}%
\definecolor{mycolor2}{rgb}{0.85000,0.32500,0.09800}%
\definecolor{mycolor3}{rgb}{0.92900,0.69400,0.12500}%
\begin{tikzpicture}[font=\footnotesize, spy using outlines={rectangle, magnification=3, size=1cm, connect spies}]
\begin{axis}[%
width=72mm,
height=38mm,
at={(0in,0in)},
scale only axis,
xmin=0,
xmax=10,
xlabel style={font=\color{white!15!black},font=\footnotesize},
xlabel={$\sqrt{\gamma}$ [m]},
ymin=-55,
ymax=-20,
ylabel style={font=\color{white!15!black},font=\footnotesize},
ylabel={Transmit power [dBm]},
axis background/.style={fill=white},
xmajorgrids,
ymajorgrids,
legend style={legend cell align=left, align=left, draw=white!15!black}
]
\addplot [color=mycolor1, line width=1.0pt, mark=diamond, mark options={solid, mycolor1}]
  table[row sep=crcr]{%
0	-20\\
0.1	-20.0000048357058\\
0.4	-20.000004467508\\
0.7	-20.0000007018878\\
1	-20.0000018114994\\
2	-20.0000022387968\\
4	-20.0000079824303\\
10	-20.0000261487422\\
};

\addplot [color=mycolor2, dashed, line width=1.0pt, mark=x, mark options={solid, mycolor2}]
  table[row sep=crcr]{%
0	-20\\
0.1	-20.0000000192803\\
0.4	-24.8730535586174\\
0.7	-29.7338145323433\\
1	-32.8318537320581\\
2	-38.8524536453378\\
4	-44.8730535586174\\
10	-52.8318537320581\\
};

\addplot [color=mycolor3, dashed, line width=1.0pt, mark=x, mark options={solid, mycolor3}]
  table[row sep=crcr]{%
0	-20\\
0.1	-20.0000001090062\\
0.4	-21.418717888898\\
0.7	-26.2794788626239\\
1	-29.3775180623388\\
2	-35.3981179756184\\
4	-41.418717888898\\
10	-49.3775180623388\\
};

\begin{scope}
    \spy[black, size=2.5cm] on (0.35, 3.88) in node [fill=none] at (4.3, 2.7);
\end{scope}

\end{axis}
\end{tikzpicture}%
    \vspace{-1.cm}
  \centerline{(b) Allowable transmit power versus $\sqrt{\gamma}$} \medskip
\end{minipage}
\vspace{-0.4cm}
\caption{Comparison between the proposed scheme and benchmarks:
(a) $\sqrt{\text{CRB}}$ at Bob versus $\sqrt{\gamma}$; (b) Allowable transmit power versus $\sqrt{\gamma}$.}
\label{compare}
\end{figure}
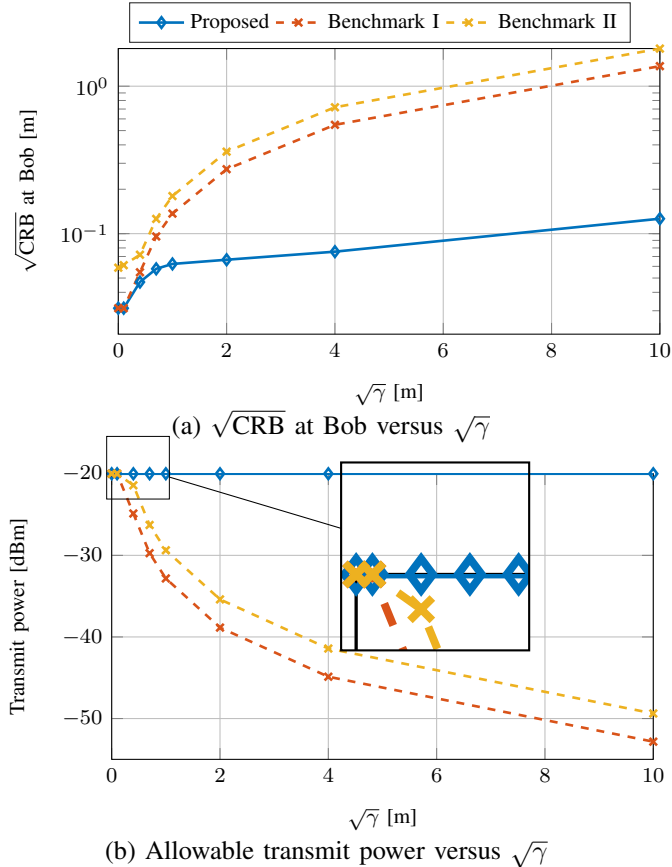

\subsubsection{Comparison Between Different Schemes}
Figures \ref{compare}(a) and (b) compare the proposed beamforming approach with benchmark methods, evaluating Bob's localization performance (characterized by the \ac{crb}) and the allowable transmit power against the requirement for location privacy protection (characterized by $\sqrt{\gamma}$), respectively. From Fig. \ref{compare}(a), we observe that the proposed scheme significantly outperforms the two benchmarks in achieving lower \ac{crb} for a given $\gamma$. Notably, when $\sqrt{\gamma} = 0\text{ m}$, Bob's \ac{crb} under the proposed scheme equals that under Benchmark I, as they are equivalent at this point. In contrast, Benchmark II results in a slightly higher \ac{crb}, as it reduces complexity at the cost of degraded localization performance\cite{furkan2022tvt}. As shown in Fig. \ref{compare}(b), the superiority of the proposed scheme arises from its ability to maintain full transmit power, while the benchmarks must reduce transmit power to satisfy the location privacy constraint (except when $\gamma$ is very small, where the privacy constraints are not restricted).
This advantage is due to the proposed scheme's ability to manage energy leakage to Eve by judiciously exploiting spatial degrees of freedom, allowing the location privacy constraint to be met without compromising transmit power, in sharp contrast to the benchmarks.

\section{Conclusion}
We examine a localization scenario involving uplink \ac{mimo}-\ac{ofdm} where a legitimate \ac{bs} sought to determine the location of a \ac{ue}, while an unauthorized \ac{bs} jeopardizes the \ac{ue}'s privacy by eavesdropping on pilot signals to estimate its position. To improve legitimate localization while safeguarding privacy, we formulate an optimization problem aimed at minimizing the \ac{crb} for legitimate localization, subject to constraints on unauthorized localization. Leveraging a \ac{pdd} framework, we propose an innovative beamforming strategy. Numerical simulations validate our approach's effectiveness, showcasing its advantages over established benchmarks.

\appendices

\bibliographystyle{IEEEtran}
\bibliography{IEEEabrv,mybib}

\end{document}